\documentclass[sigchi]{acmart}

\usepackage{booktabs} % For formal tables
\usepackage{graphics}      % for EPS, load graphicx instead 
\usepackage{color}
\usepackage{textcomp}
\usepackage{soul,color}
\usepackage{subfig}
\usepackage{capt-of}
\usepackage{balance}
\usepackage[symbol]{footmisc}
\usepackage{tikz}
\usepackage{adjustbox}

\renewcommand\footnotetextcopyrightpermission[1]{} % removes footnote with conference information in first column
\setcopyright{none}
\begin{document}
\title{Excuse me! Perception of Abrupt Direction Changes\texorpdfstring{\\}{ }Using Body Cues and Paths on Mixed Reality Avatars}
% \titlenote{Produces the permission block, and
%\subtitle{Extended Abstract}
%\subtitlenote{The full version of the author's guide is available as
 % \texttt{acmart.pdf} document}

\author{Nicholas Katzakis}
%\authornote{Dr.~Trovato insisted his name be first.}
\orcid{0000-0002-1120-5648}
\affiliation{%
  \institution{Department of Informatics\\Universit\"at Hamburg}
  %\streetaddress{Vogt-Köln-Str. 30}
  %\city{Hamburg} 
  %\state{Hamburg} 
}
\email{nicholas.katzakis@uni-hamburg.de}

\author{Frank Steinicke}
\affiliation{%
  \institution{Department of Informatics\\Universit\"at Hamburg}
}
\email{frank.steinicke@uni-hamburg.de}

% \renewcommand{\shortauthors}{N. Katzakis et al.}

\begin{teaserfigure}\centering
\includegraphics[width=0.75\textwidth]{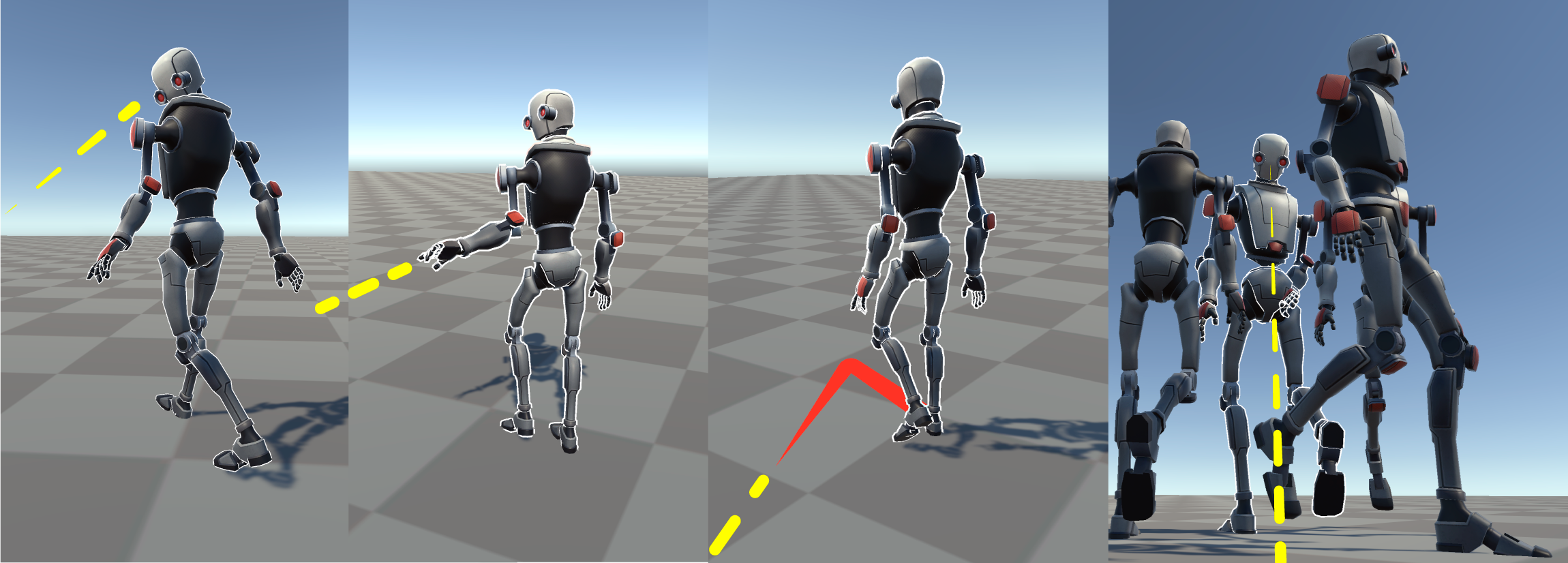}
\captionof{figure}{(1) and (2) Robot pointing using its body (3) Red path visualised on the ground (4) Snapshot from the user study with three occluding robots.
\label{fig:teaser}}
\end{teaserfigure}

\begin{abstract}
We evaluate two methods of signalling abrupt direction changes of a robotic platform using a Mixed Reality avatar. The ``Body'' method uses gaze, gesture and torso direction to point to upcoming waypoints. The ``Path'' method visualises the change in direction using an angled path on the ground. We compare these two methods using a controlled user study and show that each method has its strengths depending on the situation. Overall the ``path'' technique was slightly more accurate in communicating the direction change of the robot but participants overall preferred the ``Body'' technique.
\end{abstract}

\begin{CCSXML}
<ccs2012>
<concept>
<concept_id>10003120.10003145.10011769</concept_id>
<concept_desc>Human-centered computing~Empirical studies in visualization</concept_desc>
<concept_significance>500</concept_significance>
</concept>
<concept>
<concept_id>10003120.10003121.10003122.10003334</concept_id>
<concept_desc>Human-centered computing~User studies</concept_desc>
<concept_significance>300</concept_significance>
</concept>
<concept>
<concept_id>10010147.10010178.10010199.10010204</concept_id>
<concept_desc>Computing methodologies~Robotic planning</concept_desc>
<concept_significance>100</concept_significance>
</concept>
</ccs2012>
\end{CCSXML}

\ccsdesc[500]{Human-centered computing~Empirical studies in visualization}
\ccsdesc[300]{Human-centered computing~User studies}
\ccsdesc[100]{Computing methodologies~Robotic planning}

\keywords{Robot, Interaction, Perception, Intentions, Gait, Externalizing, AR, VR}

\maketitle

\section{Introduction}
Humans have the expectation that robotic agents behave with a degree of social intelligence~\cite{dautenhahn2004robots,kidd2005sociable} so that bystanders can interpret their state and predict their action. The diverse forms and locomotion methods~\cite{murata2002m} of modern robots, however, make it challenging to impart social characteristics. Some research to date has attempted to make robots more predictable. This was done primarily by exploring humanoid robot designs, by adapting their gestures and gaze \cite{szafir2015communicating}, and by adapting their locomotion to be more expressive and predictable~\cite{hoque2012integrated,sadrobot,breazeal2005effects}. All of these approaches however, impose a number of constraints on the design of the robot or on its locomotion method. i.e. If the robot must perform expressive movements to cater for perception from bystanders, then it cannot travel to its next waypoint using the shortest trajectory possible, or with the most energy efficient locomotion method. 

\begin{figure}[htb]
\centering
\includegraphics[width=0.8\linewidth]{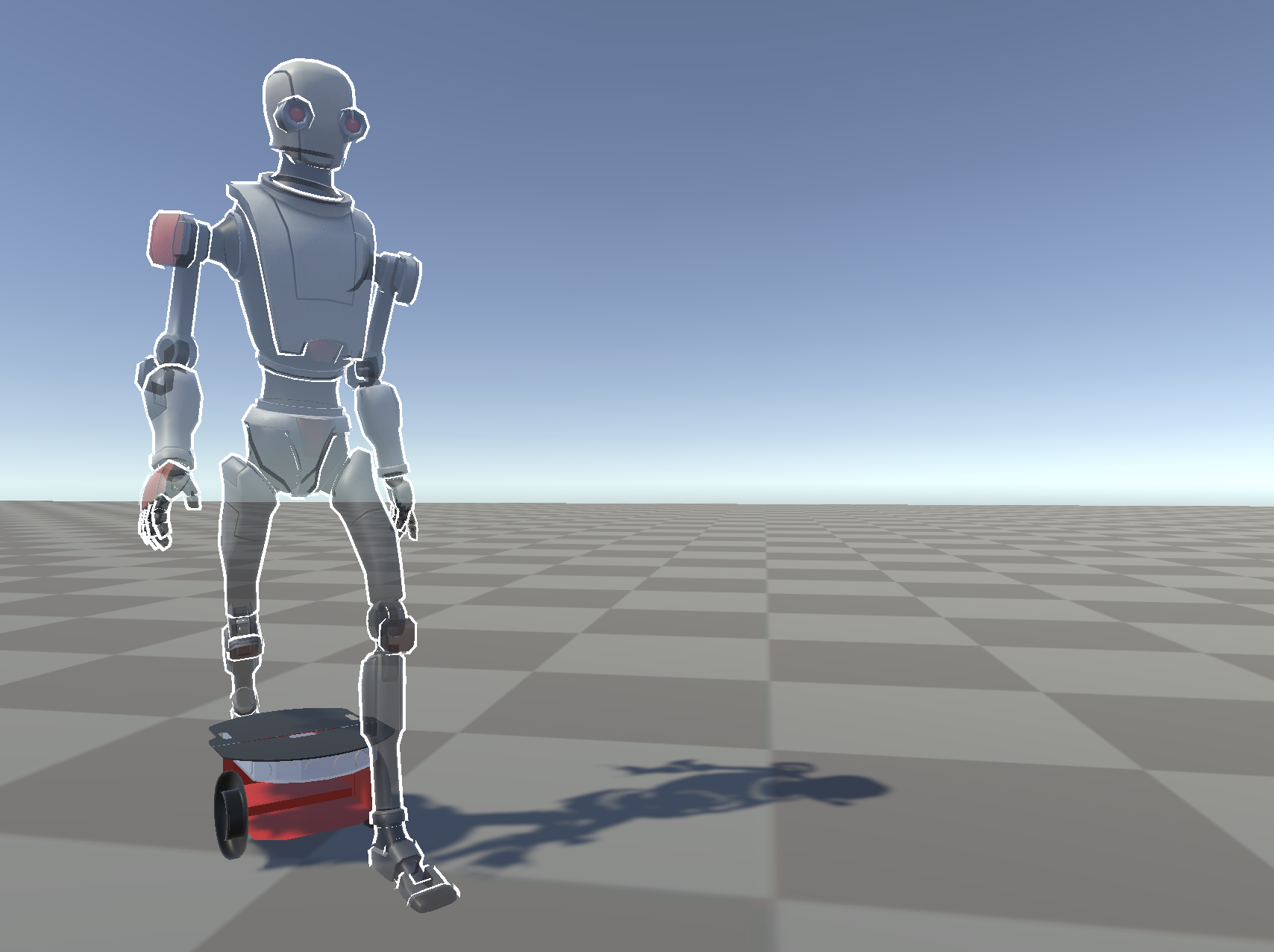}
\caption{Illustration of an envisioned application of MR agents. A humanoid avatar is superimposed on top of a real, physical wheeled robot.}
\label{fig:application}
\end{figure}

\emph{Mixed Reality} (MR) offers a possible solution to this problem. Mixed reality agents~\cite{mira} seperate the physical form of the robot from its outward appearance. The MR approach assumes that robots broadcast their future trajectories and state etc. to bystanders. Bystanders can then use their choice of MR display technology to display the robots -- and their actions -- in a way that is easier to perceive, thus enhancing their comfort in the vincinity of robots. An example of the envisioned scenario can be seen in Figure \ref{fig:application}.

This work specifically examines direction changes because they are a potential cause of discomfort to bystanders. When humans make significant changes in walking direction they convey this change with a number of body cues including gaze~\cite{hollands2002look}, slowing down, etc. Contrary to humans, robots can change the parameters of their locomotion instantly~\cite{watanabe1998feedback} to account for a new heading or to avoid obstacles. Therefore, even if bystanders trust that robots will not collide with them, these abrupt direction changes might be a source for distraction and discomfort. 

The proliferation of autonomous agents and vehicles raises the question: Is it possible, with MR agents, to warn bystanders of abrupt direction changes? We explore the design space and present two potential solutions: One uses body cues of a humanoid MR avatar, the other displays the direction change as a path on the ground (Figure \ref{fig:teaser}). The two approaches are evaluated in a controlled user study that manipulates \emph{cue onset timing}, \emph{cue expressiveness} and \emph{robustness to occlusion}. 

\section{Related Work}
Shoji et al. superimposed an avatar on a robot to increase attractiveness~\cite{shoji2006u} while Shimizu et al.~\cite{shimizumime} superimposed an avatar that attempts to mimic the movements of the robot. Aspects of intent or predictability, however, remain largely unexplored.

The approach of Dragone and O'Hare~\cite{dragone2007using,dragone2006mixing,mira} makes some headway towards overcoming these limitations. In the context of a \textbf{Mi}xed \textbf{R}eality \textbf{A}gent (MiRA), they display a humanoid avatar that appears sitting on a rhoomba robot to communicate the state of the robot. Such an avatar can indicate direction changes by pointing yet the authors have not evaluated their proposed system in this context.

Young et al. explored a similar concept using cartooning~\cite{youngCartoons}. They separated \emph{behavioural likeness} from \emph{visual likeness}, in an attempt to avoid the ``uncanny valley''~\cite{uncannyvalley}. These cartoon-like illustrations offer a simple, yet powerful means of expressing various states of the robot. It remains unclear how this approach could be applied to express spatial intent and direction changes.

The idea of externalizing the internal state of a robot using AR visualisations has been explored by Collett et al.~\cite{collett2010augmented} in the context of debugging the various sensors of the robot. More recently H{\"o}nig et al.~\cite{mrforrobotics} also used an augmented reality setup with the goal of creating a safer environment for debugging  algorithms where robots are to be used in close proximity to humans. Other works have explored path planning~\cite{zollmann2014flyar,giesler2004using,stilman2005augmented}, also in an attempt to assess accuracy and diagnose waypoint errors. 

Hoffman and Ju~\cite{hoffman2014designing} have suggested that rather than visual likeness to humanoids, motion can be a powerful tool to communicate the state of the robot. Although this approach is ideal for communication when the robot is stationary, instant waypoint change situations leave little time for expressive movements.

\section{Design}
The design goal of the two proposed methods is to inform bystanders that the robot will change its direction and allow them to predict \emph{where the robot is going next}. The assumption being that bystanders will feel more comfortable knowing where the robot is heading. Rather than superimposing a realistic human avatar, we chose the ``Kyle'' robot model in unity because of its torso. The slim torso reduces limb occlusion from various angles and the elbow and shoulder guards make arm rotations more salient.

\subsection{Body}
\begin{figure*}[htbp]
\centering
\captionsetup[subfigure]{hangindent=0pt,singlelinecheck=false}
\subfloat[][Body low expressivity: Only the head orients towards the waypoint.]{\includegraphics[width=.2\linewidth]{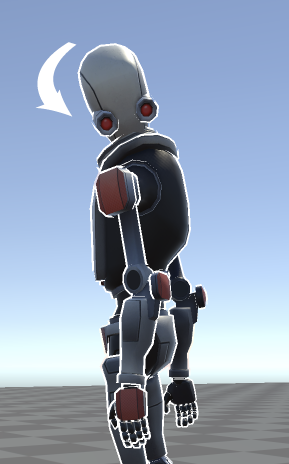}\label{fig:body-low}}
\hspace{1em}
\captionsetup[subfigure]{hangindent=0pt,singlelinecheck=false}
\subfloat[][Body medium expressivity: In addition to the head, the arm closest to the target points.]{\includegraphics[width=.2\linewidth]{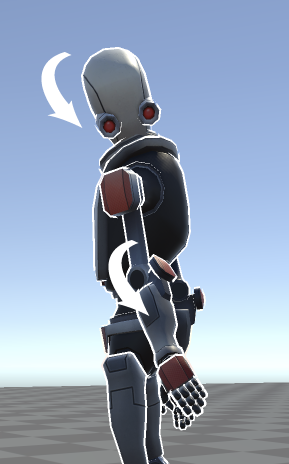}\label{fig:body-med}}
\hspace{1em}
\captionsetup[subfigure]{hangindent=0pt,singlelinecheck=false}
\subfloat[][Body high expressivity: In addition to the other two cues the entire torso completely orients towards the waypoint.]{\includegraphics[width=.2\linewidth]{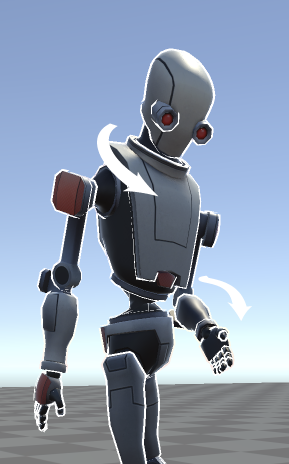}\label{fig:body-high}}
\hspace{1em}
\captionsetup[subfigure]{hangindent=0pt,singlelinecheck=false}
\subfloat[][Path low expressivity: One meter length. Point A is the start of the path; point B is the point where the waypoint change stride will occur. The other two levels are identical except with two and three meters length.]{\includegraphics[width=.2\linewidth]{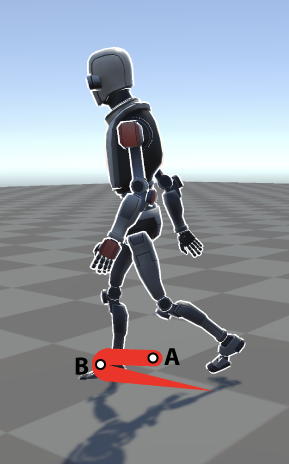}\label{fig:path-low}}
\caption{\label{fig:levels}Expressivity levels.}
\end{figure*}

When choosing cues for the \emph{body} method, we opted for cues that are easy to interpret by humans. Previous studies on human behavior in steering situations found that participants typically align their head \cite{hollands2002look,patla1997and,patla1999online} with the next waypoint immedately following delivery of a direction change cue. This head alignment occurs before torso orientation~\cite{patla1999online}. The fact that people expect such cues from other humanoids in their vincinity, inspired the use of a humanoid avatar and the first two expressivity levels (Figure \ref{fig:levels}). 

The medium expressivity level is inspired by how in some cultures people extend their forearm forward as a sign of politeness.
% \begin{figure}[htb]
% \centering
% \includegraphics[width=0.8\linewidth]{mae}
% \caption{Typical polite gesture when interrupting the path of a bystander in east asia.}
% \label{fig:shitsurei}
% \end{figure}
A third level of expressivity was added by exaggerating the turn of the avatar torso, in the hope that it will make the waypoint change cues more salient.

\subsection{Path}
Earlier works have explored visualising the path of the robot~\cite{stilman2005augmented,omidshafiei2015mar}, yet these designs extend lines far into the future and this means that in an MR setting where all the robots in the vincinity broadcast their paths the visual field would quickly become cluttered. We were therefore interested in exploring the potential of a shorter path with a pointed tip to indicate direction.

A path displayed at the feet of the avatar is possibly difficult to perceive. i.e. when a person is fixating forward while walking, the path of an adjacent robot will be outside his or her field of view. Bringing the path up to the view frustum of the person might introduce unwanted view obstruction. A further limitation of a path or arrow displayed along the line of sight is that it is difficult to judge their magnitude due to perspective~\cite{OGLE19491069}.

The displayed path begins between the feet of the robot, joins the point where the robot will change direction and extends towards the next waypoint \emph{without reaching} it (Figure \ref{fig:path-low}). The path maintains its length the entire time it is visible i.e. it extends towards the next waypoint with the same speed the robot is walking.  We manipulate the length of the path as a means of manipulating \emph{expressivity} in the \emph{path} method.

\section{Evaluation}
We conducted a controlled study to evaluate the robustness of the two proposed methods with regards to:
\begin{itemize}
	\item[$\circ$] Cue onset timing: How long do these methods need to be displayed in order to be effective in communicating the robot's waypoint?
	\item[$\circ$] Cue expressiveness: How subtle do these cues need to be?
	\item[$\circ$] Robustness to occlusion: How do these cues perform in a situation with numerous occluding robots?
\end{itemize}

Conducting this study in Virtual Reality (VR) allows for studying the perception of the robotic avatar isolated from unwanted environmental confounding factors.

\subsection{Participants - Apparatus}
\begin{figure}[htb]
\centering
\includegraphics[width=.6\linewidth]{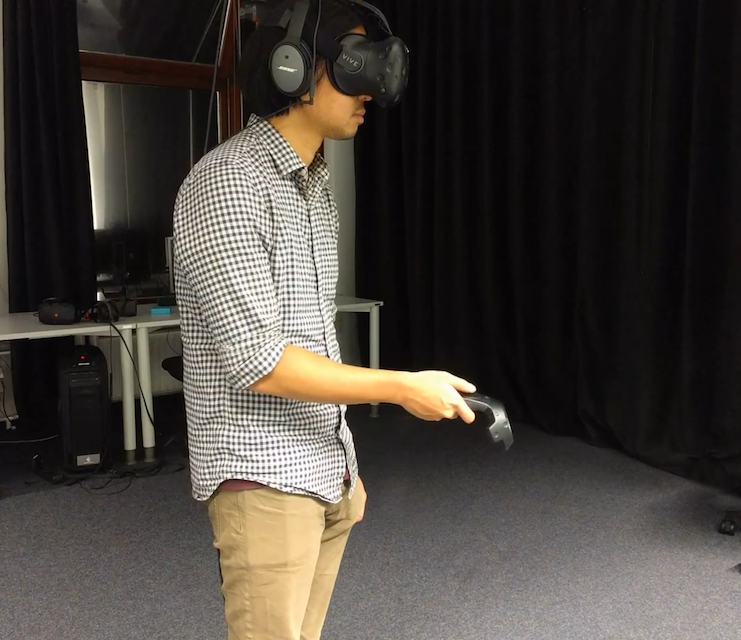}
\caption{A participant during the experiment.}
\label{fig:jon}
\end{figure}

Fourteen unpaid participants (Ages 22-38, mean age 31) were recruited from the faculty and students of <anonymized>. 75\% of participants were right handed and 25\% left handed. 91\% of participants had used a virtual reality headset before. In post-experiment questionnaires participants reported a high-level of attention during the experiment. Written and informed consent was obtained from each participant and the experiment was approved by a local ethics committe. Participants stood in the center of a 4x4 meter space and mounted the HTC Vive on their head while holding one of the Vive controllers on their dominant hand as pointer (Figure \ref{fig:jon}). Auditory cues were delivered through noise cancelling headphones. The experiment lasted approximately 40 minutes and participants were free to rest at any point during the experiment.

\subsection{Stimuli and procedure}
The virtual environment consisted of a 100 x 100 meter plain tiled floor on which participants were free to roam. At the beginning of the experiment participants were allowed to familiarize themselves with the VR environment and the controls.

\begin{figure}[htb]
\centering
\includegraphics[width=\linewidth]{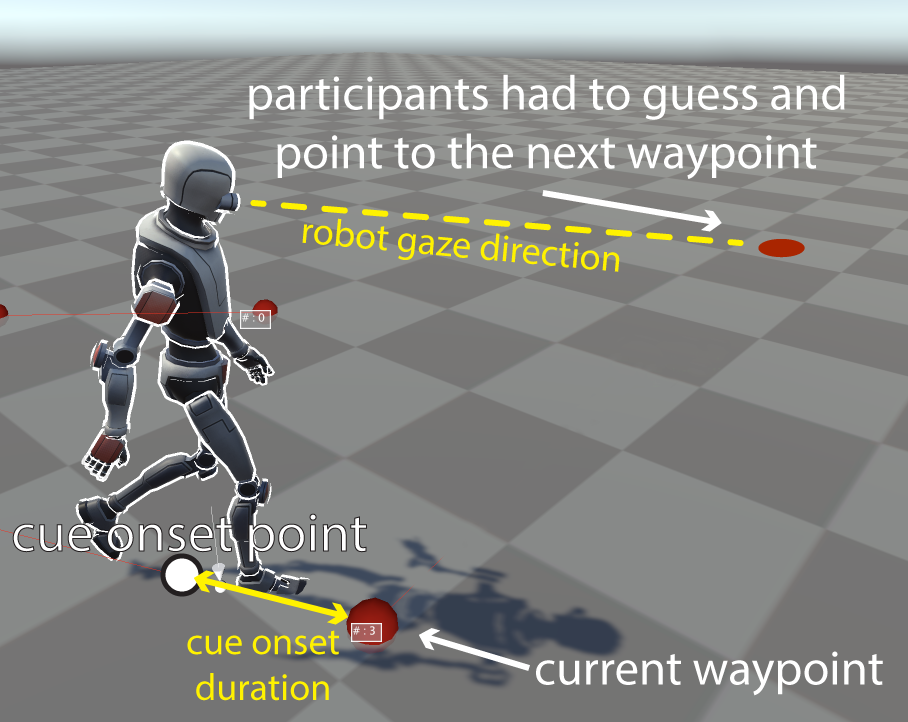}
\caption{Experiment task. Participants had to guess where is the next waypoint of the robot.}
\label{fig:guess}
\end{figure}

When the experiment started the avatar walked segments between 1.5 and 3.0 meters. At three preset onset distances, before reaching the end of these segments, the waypoint change cue would appear (body or path).  Participants were instructed to point using their wand to indicate where they thought the next waypoint of the robot will be (following the direction change -- Figure \ref{fig:guess}). Participants received visual feedback about their pointing location by a ray extending from their wand in addition to a sphere at the location where the ray intersected with the floor. Participants were free to walk around and teleport using the thumb button (pad) of the Vive controller using a standard VR teleportation technique.

The robot continued cycling waypoints with a walking speed of 1 meter/sec and participants had to keep guessing waypoints. If participants failed to indicate the correct waypoint in time the trial was queued at the end of the trials queue to be repeated later. Method was treated as one factor with two levels, \emph{Body} and \emph{Path}.In addition to the cue presentation method there were three factors with three levels each.
\begin{itemize}
\item[$\circ$] \emph{Cue onset distance}, manipulated with three levels: 0.8 m, 1m and 1.5m. In addition we manipulate
\item[$\circ$] \emph{Cue Expressivity}, manipulated with three levels: low, medium and high.
\item[$\circ$] \emph{Number of occluding robots}, manipulated with three levels: no robots, 3 robots and 6 robots.
\item[$\circ$] \emph{Angle} at which the robot turned, manipulated at 6 levels: -30,-20,-10,10,20,30.
\end{itemize}

\begin{figure}[htb]
\centering
\includegraphics[width=\linewidth]{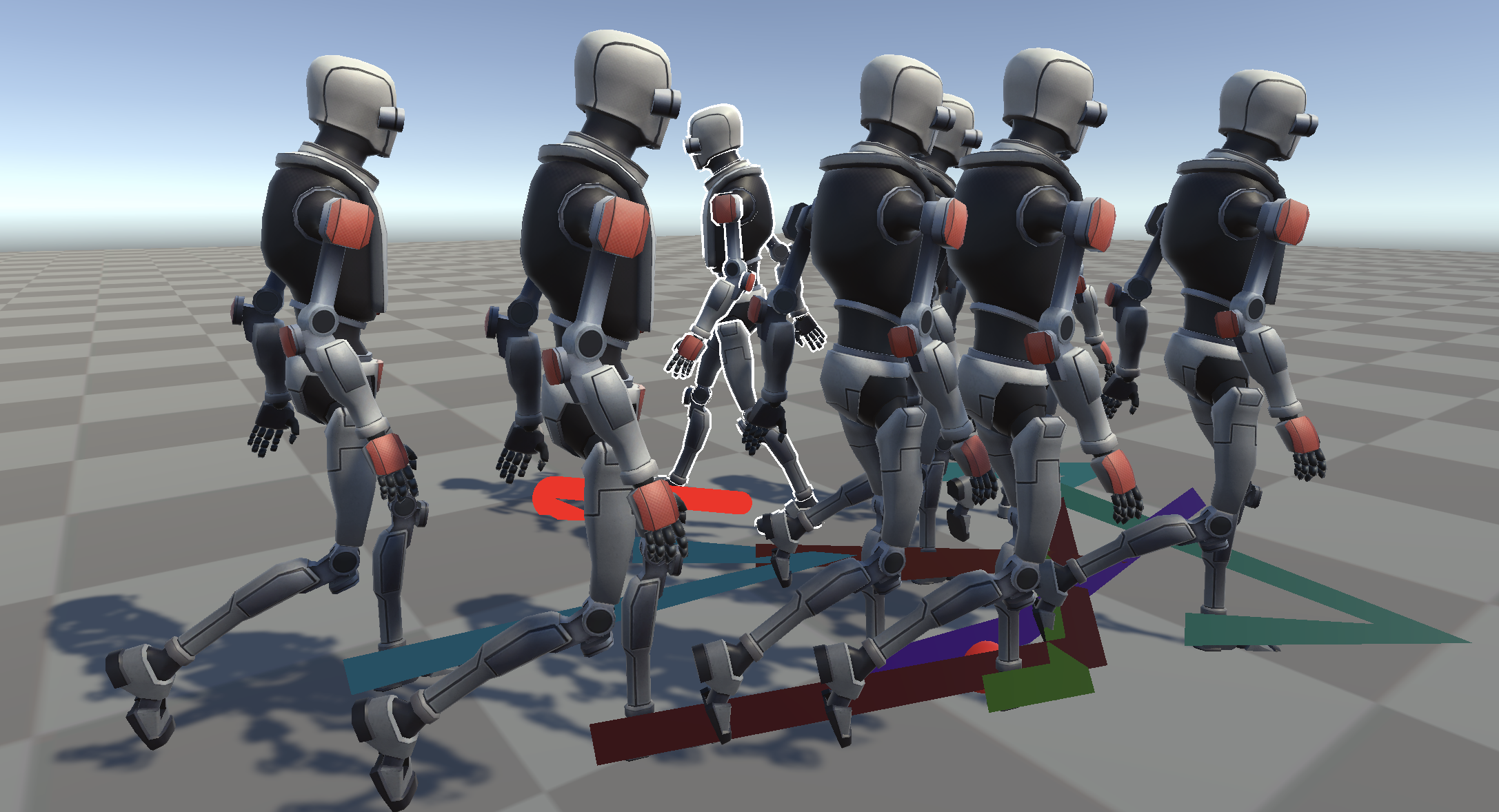}
\caption{Six Occluding robots in the \emph{Path} condition. The main robot avatar has a faint white outline for disambiguation.}
\label{fig:robots}
\end{figure}

The occluding robots spawned on a random location in a circle of 3m radius around the main robot avatar. The robots chose random waypoints in a circle with a 3m radius centered two waypoints ahead of the main robot avatar. The result was random occlusion and intersection of the main avatar's path. Rather than passing through one another without collisions, in a real situation the robots should perform collision avoidance. Preventing collisions was a conscious design decision on our part which aimed to disentangle visual cues from behavioral prediction cues formed by prior lifetime experiences of the participants. i.e. We did not want participants to be able to predict that robots will change direction to avoid collision. They should only rely on the proposed direction change cues. The main avatar had a white outline attached in all conditions so that it can be distinguished from the rest of the robots (Figure \ref{fig:robots}). 

All factors were randomized except cue method and number of occluding robots. Those were repeated in counter balanced blocks. e.g. Body method was tested with 3 distracting robots, followed by 0 robots, followed by 9 robots etc. Following one Body x Distracting Robots block the Path method was tested again with all Distracting robot levels. This was counter balanced across participants. 

This design was chosen because the experiment was too challenging if robot numbers and technique were alternated with every trial (robots appearing/disappearing etc.). Given that the robot travelled randomly between 1.5m-3m in every path segment this resulted on an average of 2.25 seconds per segment. The occluding robots travelled at the same speed (1 m/sec) as the main avatar. Because of this, if they randomly disappeared and spawned with every single trial, there would be insufficient time for them to walk around and form a random pattern while the appearance and disappearance of robots would be too distracting and detached from the envisioned application scenario of this work.

We recorded the \emph{reaction time} (RT), i.e. time between cue onset and trigger press by the participant. We also recorded the prediction accuracy of participants, by recording the \emph{Error}. i.e. The euclidean distance between the actual robot waypoint and the location the participant indicated with the wand.

The experiment was a within-subjects design with 2 methods x 3 onset distances x 3 expressivity levels x 3 occluding robot levels x 6 angles x 2 trials = 648 trials. 14 participants x 648 trials per participant = 9072 total trials collected. The data was analyzed using a repeated-measures ANOVA test in R.

\section{Results}
As expected, method had a significant effect on both Error ($F_{1,13}=20.78$, $p<0.01$) and RT ($F_{1,13}=6.12$, $p<0.01$). 

\begin{figure}[htb]
\centering
\includegraphics[width=\linewidth]{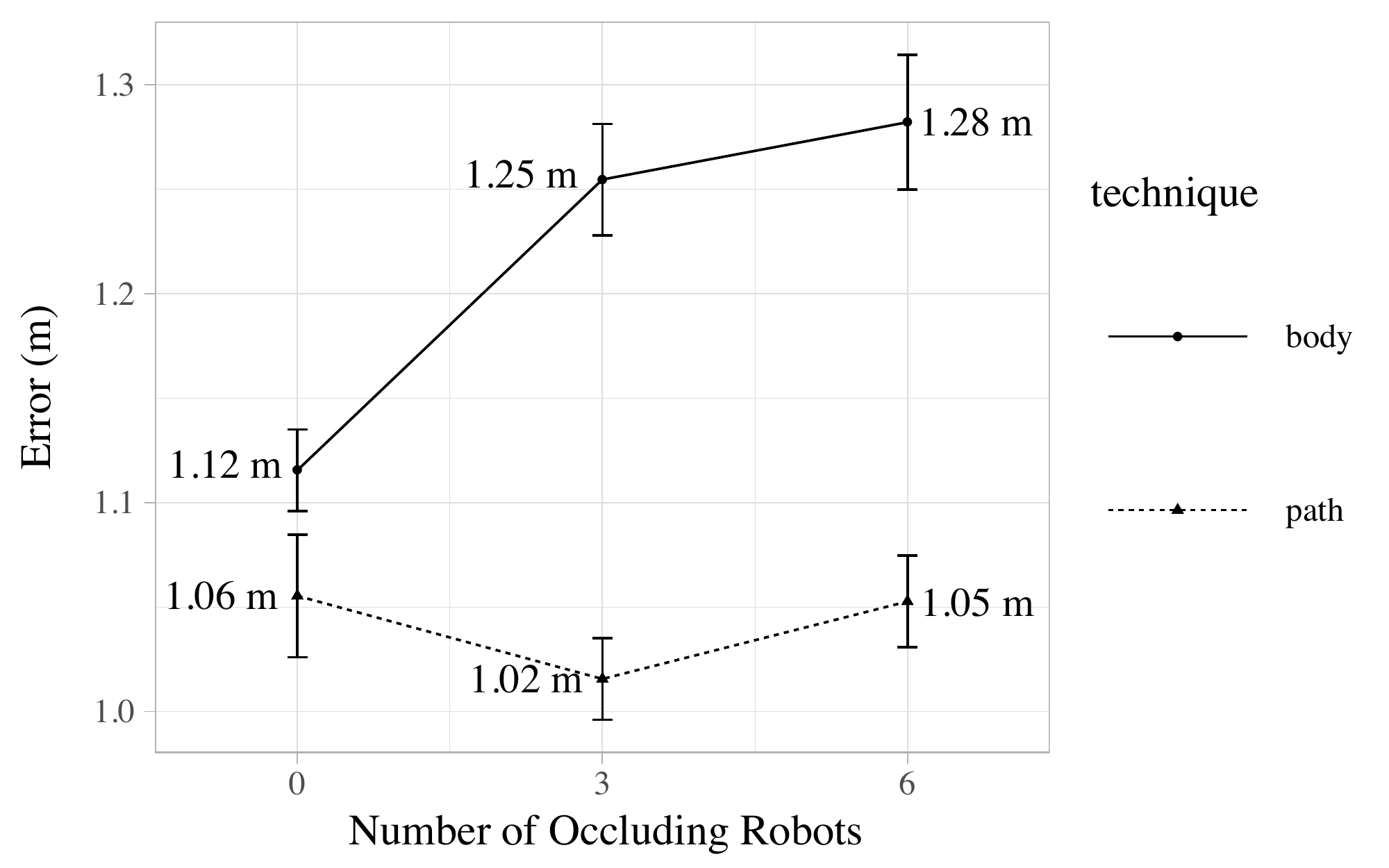}
\caption{Effect of number of distractor robots on Error (lower is better). The body method was more sensitive to the number of distractors.}
\label{fig:distractors}
\end{figure}

There was a significant effect of number of distracting robots (Figure \ref{fig:distractors}) on Error ($F_{1,13}=6.12$, $p=0.02$). Number of robots had no effect on RT ($F_{1,13}=8$, $p=0.6$).

\begin{figure}[htb]
\centering
\includegraphics[width=\linewidth]{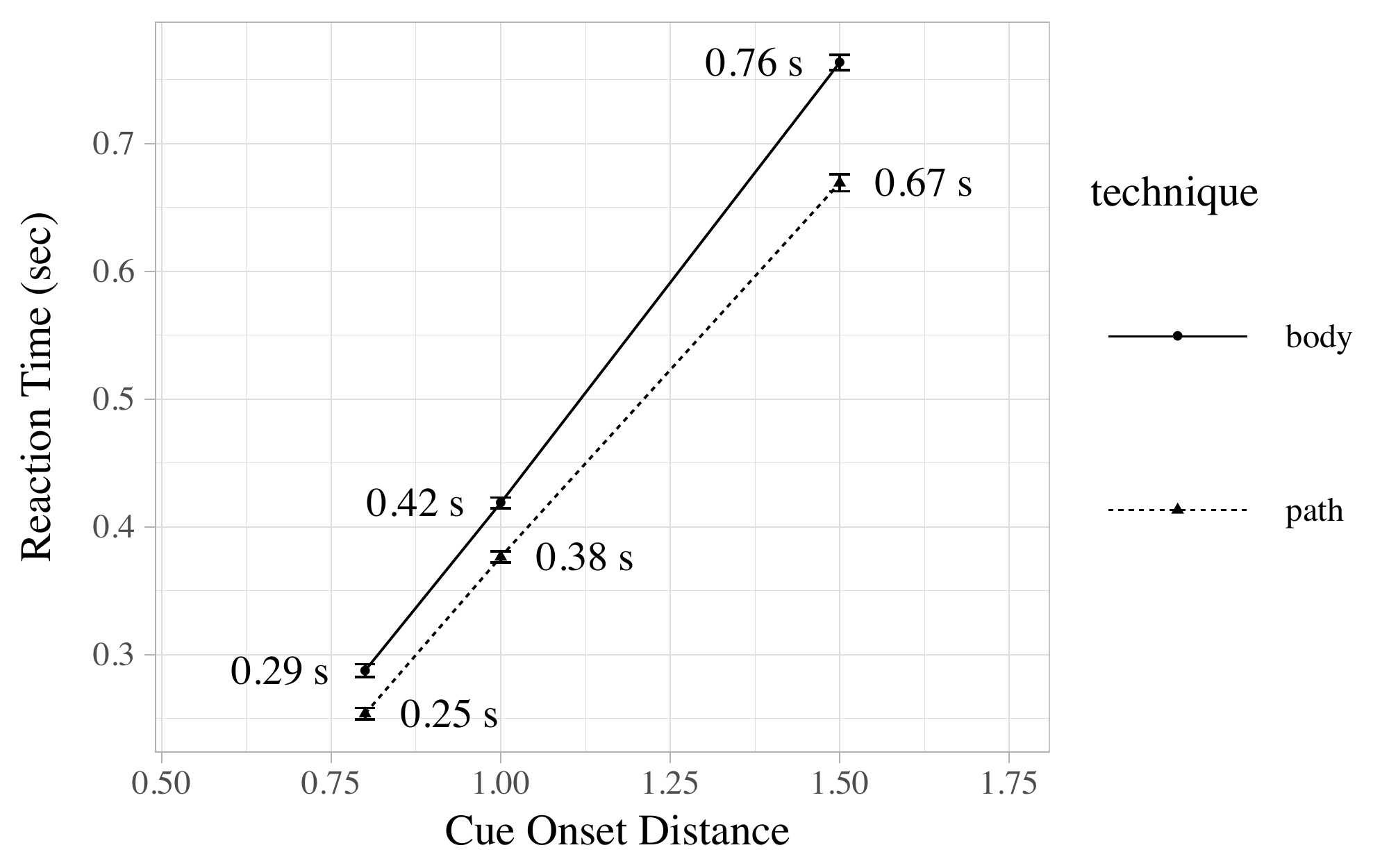}
\caption{Effect of cue onset distance on RT.}
\label{fig:cueonset_error}
\end{figure}

Cue onset distance did not have a significant effect on Error ($F_{1,13}=0.01$, $p=0.9$) but, as expected, a highly significant effect on RT ($F_{1,13}=384$, $p<0.01$). Figure \ref{fig:cueonset_error} contains a graph of cue onset distance on RT.

\begin{figure}[htb]
\centering
\includegraphics[width=\linewidth]{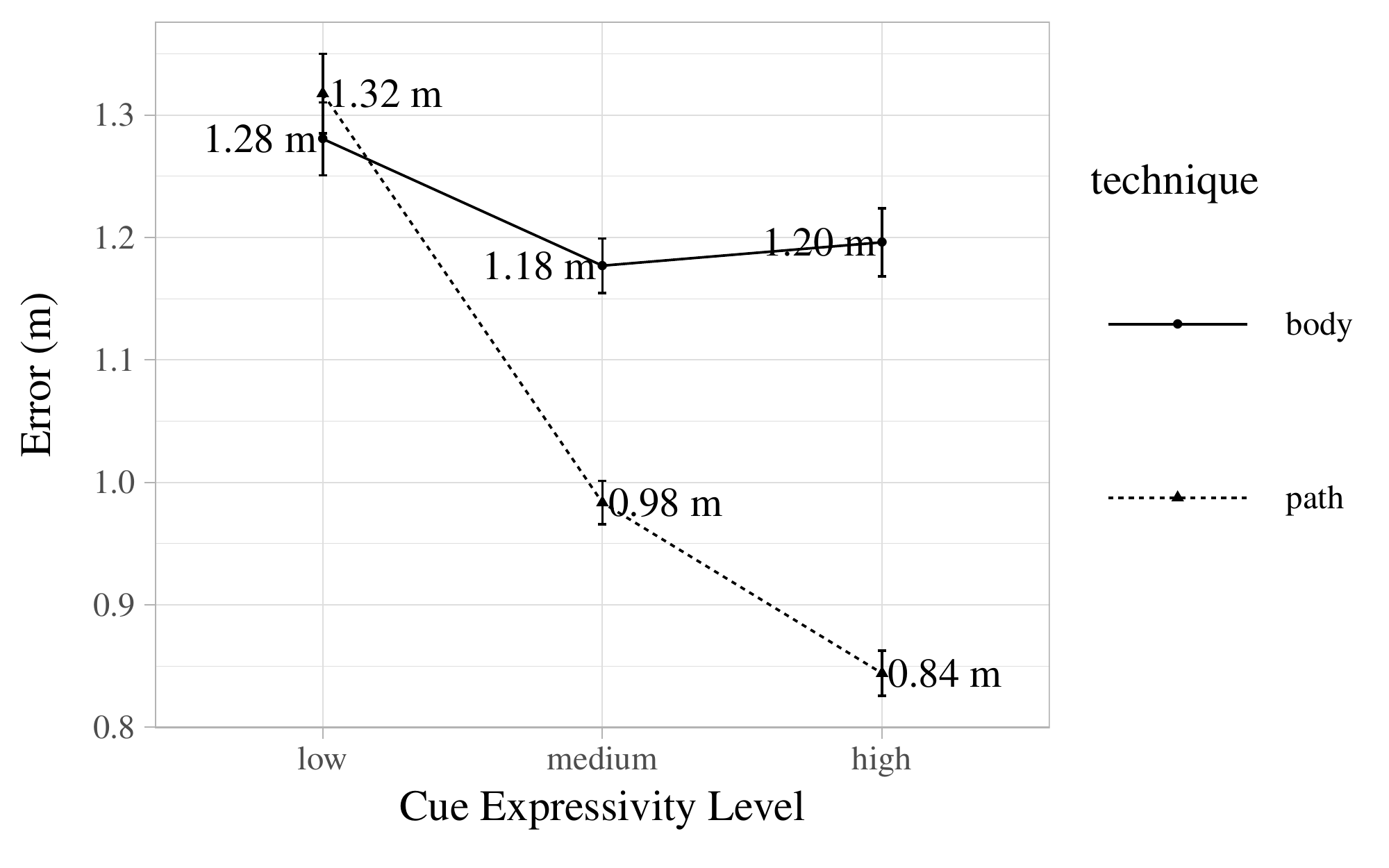}
\caption{Effect of cue expressivity level on Error.}
\label{fig:expressivity}
\end{figure}

\emph{Expressivity level} had no statistical effect on RT ($F_{2,26}=0.3$, $p=0.7$) but had a significant effect on Error ($F_{2,26}=25.5$, $p<0.001$). A graph is displayed in Figure \ref{fig:expressivity}).

\begin{figure}[htb]
\centering
\includegraphics[width=\linewidth]{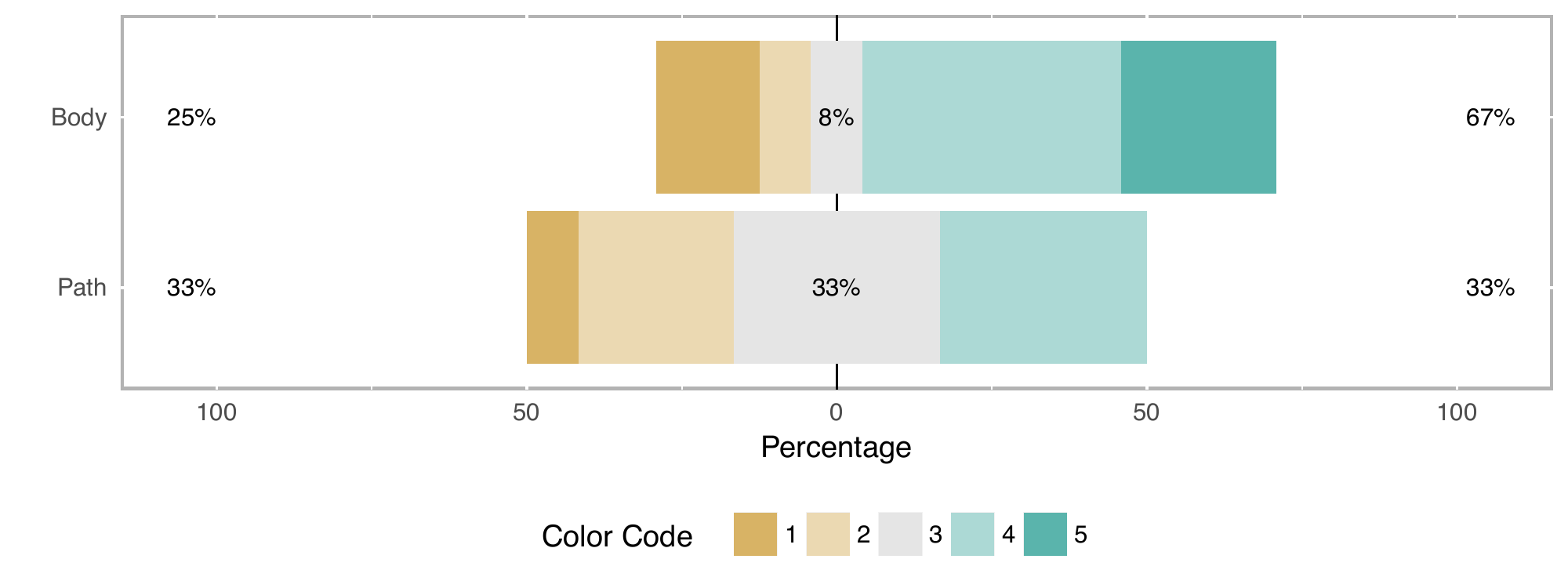}
\caption{Participant's ratings of the methods on a 5-point likert scale. The \emph{Body} method was preferred.}
\label{fig:likert}
\end{figure}

In a post-experiment questionnaire users were asked to rate both methods in terms of "Ease of Perception" on a 5-point likert scale. A summary of the results can be seen in figure \ref{fig:likert}.

There was no significant effect of \emph{Angle} on Error or RT ($F_{2,26}=1.37$, $p<0.26$).

\section{Discussion}
As expected, cueing method (Body or Path) affected the participant's ability to accurately sense prediction. Using the \emph{Body} method participants were on average 1.21m away from the robot's waypoint vs. 1.04 m from the path. A difference of approximately 17cm. On the other hand, on the low expressivity level, when the path was shortest and the avatar only turned its head, the Body technique performed significantly better ($F_{2,26}=21.3$, $p<0.001$ - Figure \ref{fig:expressivity}). Summarizing the responses from the questionnaires shows that participants preferred the \emph{Body} technique (Figure \ref{fig:likert}).

It is also notable that the length of the path (cue expressivity) resulted in a significant improvement in performance from the low to the high condition. The 3m path resulted in an average \emph{Error} of 0.84 m vs. 1.32 m for the 1m path. Despite this the change is not a multiple of the path length (non-linear) and this suggests that MR interface designers should not extend the line too far into the future of the robot's planned path. This makes additional sense when considering that paths might be updated in response to external factors.

The \emph{path} method also proved more robust to visual occlusions than the \emph{body} method. Specifically, the path technique's Error remained around 1.03 m despite the number of robots present on the scene (Figure \ref{fig:distractors}). In a busy environment with many robots present designers might use the path technique whereas when there are fewer robots they could signal direction changes using the body cues alone. This is especially important considering that when robots are near, the path might be outside the person's field of view.

Finally, cue onset distance results can be directly translated to time (in seconds) because the robot was travelling with a 1 meter/second speed. Results from our experiment suggest that the path method had a marginally shorter reaction time than the body method. This is another result that can be considered by designers when choosing the appropriate cue for the moment. Although we assume that robots would always cater to avoid collisions with humans, mechanical errors could occur. In such a situation a robot might be aware of its own malfunction and broadcast its direction change for bystanders to avoid it. When the robot is very near the participant's visor could chose the path technique when reaction time is of the essence. 

One participant commented that when there were many occlusions from the other robots, they occasionally observed the shadow to find the direction. This makes sense because an arm extended from the robot's shadow forms a perfectly aligned 2D vector on the ground, pointing to the desired waypoint and is perhaps an additional design space for this domain.

The results from this experiment suggest that each technique has its own strengths and robotics MR interface designers can take these strenghts and weaknesses into account when choosing which method to use.

A limitation of this study is that direction changes were considered only for locomotion across the ground plane. What if the robot is traveling in other axes? Can these cues be extended to convey direction changes when robots are performing actions other than walking like drones or other, more exotic robots that jump around (like a spider robot)?

\section{Conclusion - Future Work}
We have presented two methods for communicating direction changes using superimposed MR avatars. Our findings suggest that both these methods have strong points and that interface designers should consider these strenghts when choosing which technique to use for their application. 

This study is a pilot exploration of a design space that we believe offers potential for rich communication. Many interesting questions remain open for future work: Can we detach the MR avatar from the actual robot so that even though the physical robot is abruptly changing position, the MR avatar starts turning in advance? Can the MR avatar follow a curved turning path even though the physical robot turns in a sharp angle? How would such a mapping be accomplished? These remain as open questions for future work.

% \begin{acks}
% This work was partially supported by the German Research Foundation - DFG Transregio SFB 169: Cross-Modal Learning. We thank the anonymous reviewers for their valuable feedback.
% \end{acks}

% BALANCE COLUMNS
\balance{}

\bibliographystyle{ACM-Reference-Format}
\bibliography{bibliography} 

\end{document}